\documentstyle[pra,aps]{revtex}

\def\mathsymfont#1{\ifmmode{\mathchoice{%
    \mbox{\the\textfont0\ignorespaces#1}}{%
    \mbox{\the\textfont0\ignorespaces#1}}{%
    \mbox{\the\scriptfont0\ignorespaces#1}}{%
    \mbox{\the\scriptscriptfont0\ignorespaces#1}}}%
  \else{{\rm\ignorespaces#1}}\fi}

\def\C{\mathsymfont{\kern.24em
    \vrule width.02em height1.4ex depth-.05ex
    \kern-.26em C}}

\def\bm#1{\mathchoice{\mbox{\boldmath{$\displaystyle #1$}}}%
{\mbox{\boldmath{$\textstyle #1$}}}%
{\mbox{\boldmath{$\scriptstyle #1$}}}%
{\mbox{\boldmath{$\scriptscriptstyle #1$}}}}

\def\code{{\cal C}}

\newtheorem{defin}{Definition}
\newtheorem{theo}{Theorem}
\newtheorem{lemma}[theo]{Lemma}
\newenvironment{proof}{{\bf Proof:\ }}{\hfill{\rule{1ex}{1ex}}\par}

\title{A Note on Non--Additive Quantum Codes}

\author{M.~Grassl\thanks{e--mail: {\protect\tt grassl@ira.uka.de}} and
Thomas Beth}
\address{
	Institut f{\"u}r Algorithmen und Kognitive Systeme,
        Universit{\"a}t Karlsruhe, Am Fasanengarten 5, 
	D--76\,128 Karlsruhe, Germany.
}
\date{March 10, 1997}

\begin{document}
\twocolumn
\narrowtext
\maketitle
\begin{abstract}
A method to combine two quantum error--correcting codes is presented.
Even when starting with additive codes, the resulting code might be
non--additive. Furthermore, the notion of the erasure space is
introduced which gives a full characterisation of the
erasure--correcting capabilities of the codes. For the special case
that the two codes are unitary images of each other, the erasure space
and the pure erasure space of the resulting code can be calculated.
\newline
{\bf Note:} This report is preliminary, any suggestions or comments on
errors are welcome.
\end{abstract}

%==========================================================================
\section{Introduction}
%==========================================================================
%\onecolumn
%\narrowtext

In a very recent paper \cite{RHSS97}, Rains {\em et al} presented the
first non--additive quantum error--correcting code. The code was
constructed using numerical iteration to build a projector with a
given weight distribution. That code was transformed into a code with
orthogonal basis $\{|\bm{c}_0\rangle,\ldots,|\bm{c}_5\rangle\}$ where
each of the basis elements is a $((5,1,3))$ additive quantum
code. Futhermore, all of them are equivalent. For $i=1,\ldots,5$ the
basis elements $|\bm{c}_i\rangle$ can be obtained by multiplication of
$|\bm{c}_0\rangle$ with the transformation
$\tau=\openone\otimes\openone\otimes\sigma_x\otimes\sigma_x\otimes\sigma_x$
and its cyclic shifts.

This generalizes to the construction of {\em union quantum codes},
i.\,e., the code is the sum of the subspaces generated by two quantum
codes of same length. The name union is motivated by the analogue
construction for classical codes. Given two (linear) codes of equal
length, the union of the first code and a properly chosen coset of the
second yields a possibly non--linear code.

%==========================================================================
\section{Quantum Codes and Erasure Spaces}
%==========================================================================
%--------------------------------------------------------------------------
\subsection{General Codes}
%--------------------------------------------------------------------------

Following \cite{Rains96}, let $\code=((N,K,d))$ denote a quantum
error--correcting code that spans a $K$-dimensional subspace of a
$2^N$-dimensional Hilbert space. The code can correct any error that
affects less than $\frac{d}{2}$ qubits, or equivalently, as shown in
\cite{GBP97}, it can correct erasures of up to $d-1$ qubits.

Recall that necessary and sufficient conditions for a quantum code
with orthogonal basis $\{|\bm{c}_i\rangle\}$ to correct up to $d-1$
erasures are
\begin{eqnarray}
\forall i\ne j: \langle\bm{c}_i|E|\bm{c}_j\rangle&=&0\label{cond_i}\\
\forall i,j: \langle\bm{c}_i|E|\bm{c}_i\rangle&=&
  \langle\bm{c}_j|E|\bm{c}_j\rangle\label{cond_ii}
\end{eqnarray}
for any error operator $E$ of weight less than $d-1$
(cf.~\cite{Knill96} for the definition of error operators and
\cite{Rains96} for the definition of their weight).

It is sufficient to consider only operators that are of the form
$E=\sigma_{i_1}\otimes\ldots\otimes\sigma_{i_N}$ where the $\sigma_i$
are the identity operator $\openone$ or one of the Pauli matrices
$\sigma_x$, $\sigma_y$, $\sigma_z$. The set of error operators of that
type will be denoted by ${\cal EB}$ and is an algebra basis of the
operators on the $2^N$--dimensional Hilbert space.

In general, conditions (\ref{cond_i}) and (\ref{cond_ii}) may be
fulfilled for some operators of weight greater then $d-1$. This
motivates the definition of the {\em erasure space\/} of a quantum
code:
\begin{defin}\ \\
The {\em erasure space} ${\cal E}(\code)\subseteq\C^{2^N\times 2^N}$
of a quantum code $\code$ is the set of operators $E$ that fulfil
(\ref{cond_i}) and (\ref{cond_ii}).
\end{defin}

Clearly, ${\cal E}(\code)$ is closed under addition. Furthermore,
$E^\dagger\in{\cal E}(\code)$ for every $E\in{\cal E}(\code)$. Thus,
every element of a ($\C$--vector space) basis of ${\cal E}(\code)$ can
be written as Hermitian operator $E+E^\dagger$ or anti--Hermitian operator
$E-E^\dagger$. This is summarized by the following lemma:
\begin{lemma}
The {\em erasure space} ${\cal E}(\code)$ is a subspace of the $\C$
vector space $\C^{2^N\times 2^N}$. It possesses a basis consisting of
Hermitian and anti--Hermitian operators of the form $E+E^\dagger$ and
$E-E^\dagger$.
\end{lemma}
A basis of the erasure space ${\cal E}(\code)$ is called an {\em
erasure basis} ${\cal EB}(\code)$ and is denoted by ${\cal E}(\code)$.

If for a fixed set of positions any erasure can be corrected, ${\cal
EB}(\code)$ contains all error--operators $E\in{\cal EB}$ that
introduce errors at that positions. This is not the case when only
some specific erasures (e.\,g.{} phase--erasures) can be corrected.

%--------------------------------------------------------------------------
\subsection{Pure Codes}
%--------------------------------------------------------------------------
Ekert and Macchiavello \cite{EkMa96} gave the following sufficient, but not
necessary condition for codes to correct erasures $E$:
\begin{equation}\label{cond_pure}
\forall i,j:
  \langle\bm{c}_i|E|\bm{c}_j\rangle=
\left\{
\begin{array}{l@{\quad}l}
  \alpha\delta_{ij} & \mbox{for ${E=\alpha\openone}$, $\alpha\in\C$}\\ 
  0                 & \mbox{else.}
\end{array}\right.                 \label{cond_s0}
\end{equation}
Similarly to Def.~1, this defines a set of operators.
\begin{defin}\ \\
The {\em pure erasure space} ${\cal E}_{\text{\rm pure}}(\cal C)$ is
the set of operators $E$ that fulfil (\ref{cond_pure}).
\end{defin}
Again, this set is closed under addition, and since (\ref{cond_i}) and
(\ref{cond_ii}) imply (\ref{cond_pure}) we get the following:
\begin{lemma}
The pure erasure space is a subspace of the erasure space, i.\,e.,
${\cal E}_{\text{\rm pure}}(\cal C)\le{\cal E}(\cal C)$.
\end{lemma}
According to the literature, codes fulfilling (\ref{cond_pure}) for
all error--operators $E$ of weight less than $d$ are called pure or
non--degenerate $d-1$--erasure--correcting codes.

%==========================================================================
\section{Union Quantum Codes}
%==========================================================================
%--------------------------------------------------------------------------
\subsection{Definition}
%--------------------------------------------------------------------------
\begin{defin}\ \\
Let $\code^{(1)}=((N_1,K_1,d_1))$ and $\code^{(2)}=((N_2,K_2,d_2))$
be\linebreak[3] 
quantum error--correcting codes with orthogonal bases
$B^{(1)}=\{|\bm{c}_1^{(1)}\rangle,\ldots,|\bm{c}_{K_1}^{(1)}\rangle\}$
and
$B^{(2)}=\{|\bm{c}_1^{(2)}\rangle,\ldots,|\bm{c}_{K_2}^{(2)}\rangle\}$,
respectively. W.\,l.\,o.\,g. we can assume $N_1=N_2=N$ and
$\code^{(1)}\perp\code^{(2)}$.  Then the {\em union code} $\code$ is
the quantum code with basis $B=B^{(1)}\cup B^{(2)}$.
\end{defin}

Note that each of the codes $\code^{(1)}$ and $\code^{(2)}$ are
subcodes of $\code$ and they are disjoint since
$\code^{(1)}\perp\code^{(2)}$. This yields the following lemma.
\begin{lemma}
The union code $\code$ is a $((N,K,d))$ code
with $K=K_1+K_2$, $d\le\min(d_1,d_2)$, and
$\code=\code^{(1)}\oplus\code^{(2)}$. 
\end{lemma}
(Here $\oplus$ denotes the sum of complex vector spaces, and the sum
is direct since the codes are orthogonal.)

%--------------------------------------------------------------------------
\subsection{The Erasure Space of Union Codes}
%--------------------------------------------------------------------------
\subsubsection{General Case}
The union code $\code^{(1)}\oplus\code^{(2)}$ can correct all erasures
$E$ that fulfil the following conditions:
\begin{eqnarray}
\forall(\mu,i)\ne(\nu,j):
  \langle\bm{c}_i^{(\mu)}|E|\bm{c}_j^{(\nu)}\rangle&=&0\label{cond_I}\\
\forall\mu,\nu,i,j:
\langle\bm{c}_i^{(\mu)}|E|\bm{c}_i^{(\mu)}\rangle&=&
  \langle\bm{c}_j^{(\nu)}|E|\bm{c}_j^{(\nu)}\rangle\label{cond_II}
\end{eqnarray}
Since for $\mu=\nu$ these conditions are exactly those for the erasure
spaces ${\cal E}(\code^{(1)})$ and ${\cal E}(\code^{(2)})$ we have
\begin{equation}\label{sup_erasure_set}
{\cal E}(\code^{(1)}\oplus\code^{(2)})\le
  {\cal E}(\code^{(1)})\cap {\cal E}(\code^{(2)}).
\end{equation}
For $\mu\ne\nu$, we have the following additional constraints:
\begin{eqnarray}
\forall i,j:
  \langle\bm{c}_i^{(1)}|E|\bm{c}_j^{(2)}\rangle&=&0\label{cond_IaA}\\
\forall i,j:
  \langle\bm{c}_i^{(2)}|E|\bm{c}_j^{(1)}\rangle&=&0\label{cond_IaB}\\
\langle\bm{c}_1^{(1)}|E|\bm{c}_1^{(1)}\rangle&=&
  \langle\bm{c}_1^{(2)}|E|\bm{c}_1^{(2)}\rangle\label{cond_IIa}
\end{eqnarray}
(in (\ref{cond_IIa}) it is sufficient to consider only one pair with
$\mu\ne\nu$ and fixed indexes $i$ and $j$). Due to these additional
constraints, the erasure space ${\cal
E}(\code^{(1)}\oplus\code^{(2)})$ is, in general, a proper subspace of
${\cal E}(\code^{(1)})\cap {\cal E}(\code^{(2)})$.

\subsubsection{Codes of Equal Dimension}
If the codes $\code^{(1)}$ and $\code^{(2)}$ are of equal dimension,
there is a unitary transformation $U\in U(2^N)$ such that
$\code^{(2)}=U\code^{(1)}$. Then each basis element of $\code^{(2)}$
can be written as $|\bm{c}_i^{(2)}\rangle=U|\bm{c}_i^{(1)}\rangle$.
Using $\langle\bm{c}_i^{(2)}|E|\bm{c}_j^{(2)}\rangle=
\langle\bm{c}_i^{(1)}U^\dagger|E|U\bm{c}_j^{(1)}\rangle$ yields that
the error spaces are conjugated, i.\,e.,
\begin{equation}
{\cal E}(\code^{(2)})={\cal E}(U\code^{(1)})
                     =U{\cal E}(\code^{(1)}) U^\dagger.
\end{equation}
Henceforth we will drop the superscript ${}^{(1)}$. From
(\ref{sup_erasure_set}) it follows that 
\begin{equation}\label{tau_sup_erasure_set}
{\cal E}(\code\oplus U\code)\le
{\cal E}(\code)\cap U{\cal E}(\code)U^\dagger.
\end{equation}
Furthermore, conditions (\ref{cond_IaA}), (\ref{cond_IaB}), and
(\ref{cond_IIa}) read
\begin{eqnarray}
\forall i,j:
\langle\bm{c}_i|E U|\bm{c}_j\rangle&=&0\label{cond_IsA}\\
\forall i,j:
\langle\bm{c}_i|U^\dagger E|\bm{c}_j\rangle&=&0\label{cond_IsB}\\
\langle\bm{c}_1|E|\bm{c}_1\rangle&=&
  \langle\bm{c}_1| U^\dagger E U |\bm{c}_1\rangle\label{cond_IIsA}
\end{eqnarray}
Since condition (\ref{cond_IsA}) and (\ref{cond_IsB}) imply
(\ref{cond_pure}) for the operators $E'=E U$ and $E''=U^\dagger E$,
resp., we have
\begin{equation}\label{pure_sup_set}
{\cal E}(\code\oplus U\code)\le 
  {\cal E}_{\text{pure}}(\code) U^\dagger 
  \cap U {\cal E}_{\text{pure}}(\code).
\end{equation}
Combining (\ref{tau_sup_erasure_set}) and (\ref{pure_sup_set}) yields
an upper bound for the erasure space of the union code. What is more,
we can compute the erasure space of the union code:
\begin{theo}\ \\
Let ${\cal M}(|\bm{c}_1\rangle,U)$ be the vector space of operators
$E$ with
\begin{equation}\label{def_M}
\langle\bm{c}_1|E|\bm{c}_1\rangle=
\langle\bm{c}_1|U^\dagger EU|\bm{c}_1\rangle.
\end{equation}
The erasure space of the union code $\code\oplus U\code$ is given by
\begin{eqnarray}
{\cal E}(\code\oplus U\code)&=&
  {\cal E}(\code)\cap
  U {\cal E}(\code) U^\dagger\nonumber\\
&& \quad{}\cap{\cal E}_{\text{\rm pure}}(\code) U^\dagger 
          \cap U {\cal E}_{\text{\rm pure}}(\code)\nonumber\\
&& \quad{}\cap{\cal M}(|\bm{c}_1\rangle,U).\label{E_upper}
\end{eqnarray}
\end{theo}
\begin{proof}
The conjunction of the defining conditions for the spaces on the right
hand side of (\ref{E_upper}) is exactly the condition for an operator
$E$ to be in ${\cal E}(\code\oplus U\code)$.
\end{proof}
The pure erasure space of $\code\oplus U\code$ can be computed
directly from the pure erasure space of $\code$:
\begin{theo}\ \\[-1.5\baselineskip]
\begin{eqnarray}
{\cal E}(\code)_{\text{\rm pure}}\cap
  U {\cal E}(\code)_{\text{\rm pure}} U^\dagger\qquad\nonumber\\
{}\cap {\cal E}_{\text{\rm pure}}(\code) U^\dagger 
    \cap U {\cal E}_{\text{\rm pure}}(\code)
  &=&{\cal E}_{\text{\rm pure}}(\code\oplus U\code)\label{E_pure}.
\end{eqnarray}
\end{theo} 
\begin{proof}
An element $E$ of the left hand side of (\ref{E_pure}) can be
written as $E=E_1=UE_2U^\dagger=UE_3=E_4U^\dagger$ with $E_i\in{\cal
E}_{\text{pure}}(\code)$. This implies
\begin{eqnarray}
\langle\bm{c}_i|E|\bm{c}_j\rangle
   &=&\langle\bm{c}_i|E_1|\bm{c}_j\rangle=0,\label{cond_E_pure}\\
\langle\bm{c}_i|U^\dagger EU|\bm{c}_j\rangle
   &=&\langle\bm{c}_i|E_2|\bm{c}_j\rangle=0,\label{Ucond_E_pureU}\\
\langle\bm{c}_i|U^\dagger E|\bm{c}_j\rangle
   &=&\langle\bm{c}_i|E_3|\bm{c}_j\rangle=0,
       \quad\mbox{and}\label{Ucond_E_pure}\\
\langle\bm{c}_i|EU|\bm{c}_j\rangle
   &=&\langle\bm{c}_i|E_4|\bm{c}_j\rangle=0\label{cond_E_pureU}
\end{eqnarray}
which proofs that $E\in{\cal E}_{\text{\rm pure}}(\code\oplus
U\code)$. Furthermore, conditions (\ref{cond_E_pure}) --
(\ref{cond_E_pureU}) are exactly those for the spaces on the left hand
side of (\ref{E_pure}) which proofs equality.
\end{proof}
Theorems 4 and 5 characterize the erasure space and the pure erasure
space as intersection of several spaces. The weight structure of these
spaces yields bounds on the weight of the erasures that can be
corrected by the union code.

\subsubsection{Equivalent Codes}
Now consider the case when the codes $\code^{(1)}$ and $\code^{(2)}$
are {\em locally permutation equivalent}, i.\,e.,
$\code^{(2)}=\tau\code^{(1)}$ with $\tau=\pi T$ where $t$ is a local
unitary transformation $T\in U(2)^{\otimes N}$ and $\pi$ is a
permutation of the qubits. In that case, conjugation with $\tau$ is
weight preserving whereas left and right multiplications, in general,
change the weight. Thus, the minimum distance of the union code is
mainly determined by the weight structure of $\tau{\cal
E}_{\text{pure}}(\code)$ and ${\cal
E}_{\text{pure}}(\code)\tau^\dagger$.

%==========================================================================
\section{Examples}
%==========================================================================
%--------------------------------------------------------------------------
\subsection{The Code of Rains {\em et al}}
%--------------------------------------------------------------------------

As mentioned before, the code of Rains {\em et al} can be constructed
as a union code of six subcodes which are $\code^{(0)}$ generated by 
$$
|\bm{c}\rangle
=|00000\rangle
 -(|00011\rangle)_{\text{cyc}}
 +(|00101\rangle)_{\text{cyc}}
 -(|01111\rangle)_{\text{cyc}}
$$
and $\code^{(i)}=\pi^i\tau\code^{(0)}$ where
$\tau=\openone\otimes\openone\otimes\sigma_x\otimes\sigma_x\otimes\sigma_x$
and $\pi$ is a cyclic shift of the five qubits.  The pure erasure
space ${\cal E}_{\text{pure}}(\code^{(0)})$ contains all
error--operators of weight one and two. There are twenty
error--operators of weight three
($E_1=\openone\otimes\openone\otimes\sigma_y\otimes\sigma_z\otimes\sigma_y$
,
$E_2=\openone\otimes\sigma_z\otimes\openone\otimes\sigma_x\otimes\sigma_x$
and their cyclic shifts) that are not in the pure erasure space.

The weight of the operators $\pi^i\tau\pi^j E_1$ is at least three
whereas the operators $\pi^i\tau\pi^j E_1$ include the operators
$$
\begin{array}{c*4{@{{}\otimes{}}c}@{\qquad}c*4{@{{}\otimes{}}c}}
\sigma_x & \sigma_z & \openone & \openone & \openone,&
\sigma_z & \sigma_x & \openone & \openone & \openone,
\end{array}
$$
and their cyclic shifts of weight two. Indeed these operators are not
in ${\cal E}_{\text{pure}}(\code)\tau$ and according to
(\ref{E_upper}) not in ${\cal E}(\code)$. Thus, the union code has
minimum distance less than three. The minimum distance is two since
${\cal E}(\code)$ contains all error--operators of weight one and
there twenty error--operators of weight two that are not in ${\cal
E}(\code)$:
$$
\begin{array}{c*4{@{{}\otimes{}}c}@{\qquad}c*4{@{{}\otimes{}}c}}
\sigma_x & \sigma_z & \openone & \openone & \openone,&
\sigma_z & \sigma_x & \openone & \openone & \openone,\\
\sigma_z & \openone & \sigma_y & \openone & \openone,&
\sigma_y & \openone & \sigma_z & \openone & \openone,
\end{array}
$$
and their cyclic shifts.

%--------------------------------------------------------------------------
\subsection{A Special Erasure Code}
%--------------------------------------------------------------------------
We start with the $((4,4,2))$--code presented in \cite{GBP97} that can
corrrect one general erasure. This code $\code^{(1)}$ is generated by 
$$
\begin{array}{rcl}
|\bm{c}_0\rangle&=&|0000\rangle+|1111\rangle\\
|\bm{c}_1\rangle&=&|0110\rangle+|1001\rangle\\
|\bm{c}_2\rangle&=&|0101\rangle+|1010\rangle\\
|\bm{c}_3\rangle&=&|1100\rangle+|0011\rangle
\end{array}
$$
The local transformation
$\tau=\openone\otimes\openone\otimes\openone\otimes\sigma_y$ yields an
equivalent code $\code^{(2)}=\tau\code^{(1)}$ generated by
$$
\begin{array}{rcl}
|\bm{c}_4\rangle&=&|0001\rangle-|1110\rangle\\
|\bm{c}_5\rangle&=&|0010\rangle-|1101\rangle\\
|\bm{c}_6\rangle&=&|0100\rangle-|1011\rangle\\
|\bm{c}_7\rangle&=&|1000\rangle-|0111\rangle.
\end{array}
$$
The union code
$\code^{(1)}\oplus\code^{(2)}=\code^{(1)}\oplus\tau\code^{(1)}$ is a
((4,8,1)) code. It has minimum distance one and cannot correct a
general erasure, but it can correct all phase--only or amplitude--only
erasures. The reason for this is that ${\cal
E}(\code^{(1)}\oplus\tau\code^{(1)})$ contains all error--operators of
weight one of the form
$\sigma_x\otimes\openone\otimes\openone\otimes\openone$ and
$\sigma_z\otimes\openone\otimes\openone\otimes\openone$ and their
cyclic shifts.

%==========================================================================
\section{Conclusion}
%==========================================================================
Motivated by the classical analogue to join the sets of codewords of
two codes to obtain a new one, the construction of union quantum codes
has been introduced simultaneously encompassing special examples of
known and new code. The the new notion of erasure spaces gives more
insight in the erasure--correcting --- and thereby error--correcting
--- capabilities of a code than just the minimum distance. The erasure
space also provides a means to compare non--equivalent codes with the
same weight distribution.

%==========================================================================
\section{Acknowledgements}
%==========================================================================

The author acknowledges fruitful discussions with J\"orn
M\"uller--Quade and Martin R\"otteler.

\end{document}